# The Role of Stem Noise in Visual Perception and Image Quality Measurement

Arash Ashtari
*Department of Electrical and Electronics Engineering, Shiraz University of Technology, Shiraz, Iran*
*E-mail: a.ashtari@sutech.ac.ir*

*Abstract*—This paper considers (reference free) quality assessment of distorted and noisy images. Specifically, it considers the first and second order statistics of stem noise – that can be evaluated given any image. In the research field of Image quality Assessment (IQA), the stem noise is defined as the input of an Auto-Regressive (AR) process, from which a low-energy and de-correlated version of the image can be recovered. To estimate the AR model parameters and associated stem noise energy, the Yule-walker equations are used such that the accompanying Auto-Correlation Function (ACF) coefficients can be treated as model parameters for image reconstruction. To characterize systematic signal-dependent and signal-independent distortions, the mean and variance of stem noise can be evaluated over the image. Crucially, this paper shows that these statistics have a predictive validity in relation to human ratings of image quality. Furthermore, under certain kinds of image distortion, stem noise statistics show very significant correlations with established measures of image quality.

## I. Introduction

THERE are various image distortions from different sources that affect visual perception. Some image distortions – like blurriness and blockiness – are signal-dependent and others, like white noise are signal-independent. Several useful approaches have been introduced to explain image distortion and assess the quality of an image. The majority employ feature extraction and training phases to describe image distortions. For instance, BLIINDS [1], BRISQUE [2], DIIVINE [3], C-DIIVINE [4] use the parameters of Generalized Gaussian Distributions (GGD) fitted to Natural Scene Statistics (NSS). Shape, mean, left variance and right variance of the ensuing distributions are among the parameters that are thought to capture the characteristics of image distortions in different domains such as spatial, DCT and wavelet decompositions. The current work takes a complementary approach and considers the "noise" as a potentially useful signature of distortion as gauged by visual perception of natural images. In particular, in the bridges the notion of a generative model that plays a key role in visual neuroscience.

Recently, psycho-visual quality metrics, inspired by human brain hypotheses for perception, have been introduced to the literature [5], [6], [7], [8]. The stem noise used in this work is motivated by appealing to the principle of variational free energy minimization in generative modeling [9,10,11,12].

In neurobiology, this provides a formal account of perception in the human brain [13]. In biology, computational protein folding follows a way through which conformational states with lower free energies are occupied [14]. In image analysis and computer graphics, it provides a principled way to infer or reconstruct the original causes of noisy and distorted images. In brief, the free energy principle entails the minimization of a variational bound on the model evidence – or marginal likelihood – of an image, under a generative model of that image. In this work, we consider a generative model based upon a spatial autoregressive process within "blocks" or "cliques" of a two-dimensional image.

The free energy can be evaluated using a (generative) model and a (sensory) data. During visual perception, the generative model can actively encode visual scenes and images in terms of the underlying causes or latent features. While state-of-the art, free-energy based quality assessment methods such as [5], [6], [7] and [8] employ changing the systems parameters; this work considers estimates of uncertainty or precision that are part of the generative model, from which a low-energy and de-correlated version of the image can be recovered. In the visual neurosciences, this aspect of perceptual inference underwrites things like salience and visual attention – and has a key aspect of veridical image reconstruction in the brain.

Understanding how our brain perceives natural images and visual scenes in the everyday life – rather than just using feature extraction, brute deep learning techniques – may therefore be important. A recent and clear explanation of requisite energy computations has been described by [15], which introduces the "stem noise" inspired by the stem cells concept in medical science that is convertible to other cell types.

In this paper, we define the framework of stem noise and consider different image distortions such as blurriness, white noise, JPEG2000, fast fading and JPEG; as well as high quality reference images. This framework provides a straightforward generative model that enables one to estimate model parameters; namely, autoregression coefficients and the amplitude of random variables (i.e., innovations of an autoregressive process). Making use of such a model, a low-energy, de-correlated version of image is reconstructed. In this setting, under some simplifying assumptions, the variational free energy can be reduced to the marginal likelihood, as evaluated by the sum of squared stem noise (i.e., the energy of the residuals of a spatial autoregressive model).

In what follows, this approximation to model evidence (i.e., marginal likelihood) was evaluated within blocks of an image. The ensuing measure of image quality was averaged over blocks to provide a candidate measure of image quality. In addition, the between-block variance was evaluated; in terms of the second order moment of stem noise. We will see below that this (between-block) average and variance has predictive validity in relation to quantitative human assessments of image quality. Specifically, different statistics have a greater predictive validity depending upon whether the image was blurred or contaminated with high-frequency (white) spatial noise. Thus, it is possible that the brain uses similar measures of image quality to assess the precision or salience of various parts of an image. In neurobiology, this would involve the computation of stem noise energy through the local horizontal or lateral connections within early visual cortex. Interestingly, this sort of computation may underlie the construction of visual salience maps; namely, regions of high salience in visual space that attract saccadic eye movements [16].

In order to show the utility of stem noise energy in characterizing the visual perception of distorted images (i.e., establish construct validity); the statistics of stem noise energy were compared with other statistics of AR model parameters. Nine well-known natural image databases were used for this assessment; such we could also consider databases containing images simultaneously contaminated with several distortions.

The remainder of this paper is organized as follows:
The proposed method is explained in Section II. Results and their interpretation are provided in Section III. Finally, conclusions are drawn in Section IV.

## II. PROPOSED APPROACH

A block diagram of the proposed approach is shown in Fig.1. First, an energy decreasing operation is applied to the image as a pre-processing step.

This preprocessing reproduces a generic and ubiquitous aspect of free energy minimization, i.e., the repetition suppression phenomenon, where evoked responses in early (lower) visual areas of the brain are reduced for predictable, relative to unpredictable stimuli [13]. Thus, a non-linear operation in Eq.1 is applied to the image. This operation Gaussianizes the probability density functions of adjacent pixels products [2]. This is achieved by removing the local mean of luminance coefficients and normalizing the natural images. This operation mimics contrast-gain masking (in early human vision [17, 18]. It is interesting to note that successful IQA metrics – like structural similarity (SSIM) [19] and visual information fidelity (VIF) [20] – benefit from such normalization operations [21].

If the luminance component of the input image with the size of $M \times N$ pixels is $x(u,v)$, then its normalized form is:

$$\hat{x}(u,v) = \frac{x(u,v) - \mu(u,v)}{\sigma(u,v) + c} \quad (1)$$

where $u \in 1, 2 \ldots M, v \in 1, 2 \ldots N$ are spatial indices. $\mu$ and $\sigma$ are the mean and standard deviation of the image defined as follows. $c = 1$ is a constant to prevent instabilities.

$$\mu(u,v) = \sum_{k=-K}^{K} \sum_{l=-L}^{L} w_{k,l}\, x_{k,l}(u,v) \quad (2)$$

$$\sigma(u,v) = \sqrt{\sum_{k=-K}^{K} \sum_{l=-L}^{L} w_{k,l} \left(x_{k,l}(u,v) - \mu(u,v)\right)^2} \quad (3)$$

Where $w = \{w_{k,l} | k = -K, \ldots K, l = -L, \ldots, L\}$ is a low-pass filter. In the proposed method, with $K = L = 1$ a 3×3 window is used for filtering.

In the second step of the proposed approach, the low energy version of the input image, $\hat{x}$, is auto-regressively modeled. In the definitions of stochastic processes, a regular process is linearly equivalent with a white noise process. A regular process could therefore be represented as the response of a minimum-phase system. By definition, a system is minimum-phase if the system and its inverse are causal and their impulse responses have finite energy [22]. See Fig.2. This means that block of an image can be modeled with a particular AR process, in which the input is a white noise process. Such a model could be considered as the kind of generative model that our brain employs to encode early visual input. In order to estimate the AR parameters, we are interested in using a model which can be converted or estimated efficiently.

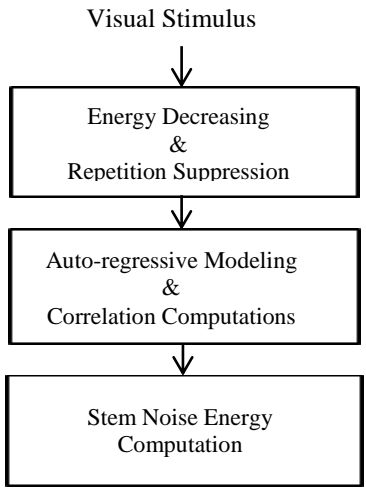

Fig.1 A block diagram of the proposed approach

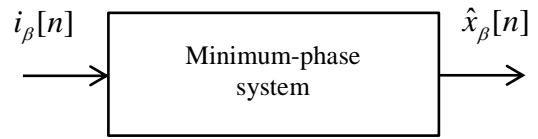

Fig. 2 A minimum-phase system with the white noise input

In the case of $r^{th}$-order AR modeling, numerical analyses suggest a reasonably optimal number of non-overlapped square blocks is:

$$s = \left\lfloor \frac{M}{\sqrt{r+1}} \right\rfloor \times \left\lfloor \frac{N}{\sqrt{r+1}} \right\rfloor \quad (4)$$

where $M \times N$ pixels is the size of the $\hat{x}$ as the normalized form of the input image and $\lfloor . \rfloor$ is the floor function and s is the number of non-overlapped square blocks. Each image block is indicated as $\hat{x}_\beta[n]$ in which $n$ is the index of image sequence indicating block pixels and $\beta$ varies from 1 to $s$. $i_\beta$ is the white noise input corresponding to the $\beta$ th block of the image. The particular $r^{th}$-order AR model for each image block is constructed as:

$$\hat{x}_\beta[n] + \sum_{m=1}^{r} a_{m\beta} \hat{x}_\beta[n-m] = b_{0\beta} i_\beta[n] \quad (5)$$

Where $a_{m\beta}$ and $b_{0\beta}$ are AR model parameters for the $\beta$ th block of the image. $b_{0\beta} i_\beta$ is the so-called "stem noise" for the $\beta$ th block. Again $\beta$ varies from 1 to $s$. Note that in this implementation of AR modeling, $\hat{x}_\beta$ is transformed to a 1-D image sequence. This is implemented by top-down row scanning, during which each row is scanned from left to right. Thus, the rightmost coefficient in the last row is referred to as $\hat{x}_\beta[n]$ and the leftmost coefficient in the first row is referred to as $\hat{x}_\beta[n-r]$. For a third-order AR model i.e. a 2×2 pixels block, the ensuing transformation is shown in Fig. 3. Since AR models could be considered as an instance of Markovian processes, it would have been possible to interpret image blocks as local cliques of a Markov random field. Considering the locality of image distortions and avoiding computational complexity [15], in this paper a third-order AR model is used for each image block. In this instance, the system function for the $\beta$ th block of the image is as follows:

$$L_\beta(Z) = \frac{b_{0\beta}}{1 + a_{1\beta} z^{-1} + a_{2\beta} z^{-2} + a_{3\beta} z^{-3}} \quad (6)$$

Where $z$ is, in general, a complex number and $\beta$ varies from 1 to $s$.

Inversion of the implicit generative model corresponds to determining the AR coefficients i.e. $a_{1\beta}, a_{2\beta}, a_{3\beta}$ and $b_{0\beta}$. This inversion problem can be solved efficiently using the Yule-Walker equations as follows.

With multiplying both sides of (5) by $i[n]$, $\hat{x}_\beta[n]$ and lags of $\hat{x}_\beta[n]$ up to $\hat{x}_\beta[n-3]$ and then taking an expectation, The Yule-Walker equations obtain as follows:

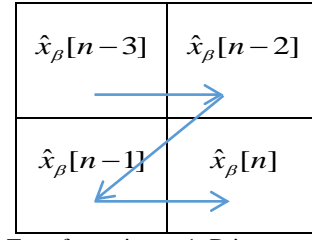

Fig3. Transformation to 1–D image sequence

$$R_{\hat{x}_\beta}(0) + a_{1\beta} R_{\hat{x}_\beta}(1) + a_{2\beta} R_{\hat{x}_\beta}(2) + a_{3\beta} R_{\hat{x}_\beta}(3) = b_{0\beta}^2 \quad (7)$$
$$R_{\hat{x}_\beta}(1) + a_{1\beta} R_{\hat{x}_\beta}(0) + a_{2\beta} R_{\hat{x}_\beta}(1) + a_{3\beta} R_{\hat{x}_\beta}(2) = 0 \quad (8)$$
$$R_{\hat{x}_\beta}(2) + a_{1\beta} R_{\hat{x}_\beta}(1) + a_{2\beta} R_{\hat{x}_\beta}(0) + a_{3\beta} R_{\hat{x}_\beta}(1) = 0 \quad (9)$$
$$R_{\hat{x}_\beta}(3) + a_{1\beta} R_{\hat{x}_\beta}(2) + a_{2\beta} R_{\hat{x}_\beta}(1) + a_{3\beta} R_{\hat{x}_\beta}(0) = 0 \quad (10)$$

where $R_{\hat{x}_\beta}$ is the auto-correlation function of the image sequence. Since in (7)-(10) equations, the auto-correlation function coefficients constitute a Toeplitz matrix, which means the equation can be solved using the Durbin-Levinson algorithm. Accordingly, the Yule-Walker equations include the Auto-Correlation function (ACF) coefficients of the natural images as the parameters of an Auto-Regressive (AR) model of the image. As illustrated in Fig.4, three model parameterizations are tied together within the Yule-Walker equations. The AR space, the ACF space, and the stem noise space. In order to solve the Yule-Walker equations, the relevant auto-correlation function can be estimated. To this end, the statistics of the adjacent pixels are used. In this paper, equation (11) is used to estimate correlation coefficients.

$$R_{\hat{x}_\beta}(p) = mean\left(\sum_{m=0}^{3-p} \hat{x}_\beta(n-m) \times \hat{x}_\beta(n-m-p)\right),$$
$$for\ p = 0,1,2,3 \quad (11)$$

To preclude conditional dependencies between auto correlation function coefficients, the term $\hat{x}_\beta(n-1) \times \hat{x}_\beta(n-2)$ is excluded from calculation of $R_{\hat{x}_\beta}(1)$.

In this case, equation (11) yields $R_{\hat{x}_\beta}(1), R_{\hat{x}_\beta}(2)$ and $R_{\hat{x}_\beta}(3)$ As parameterizing horizontal, vertical and diagonal correlations of the distorted image, respectively. After evaluating AR model parameters, the extent of implicit noise energy in each block of the input image is estimated.

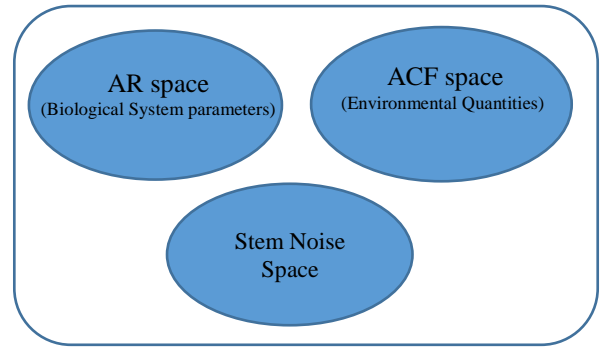

Fig 4. Different spaces are tied together in the Yule-Walker equations

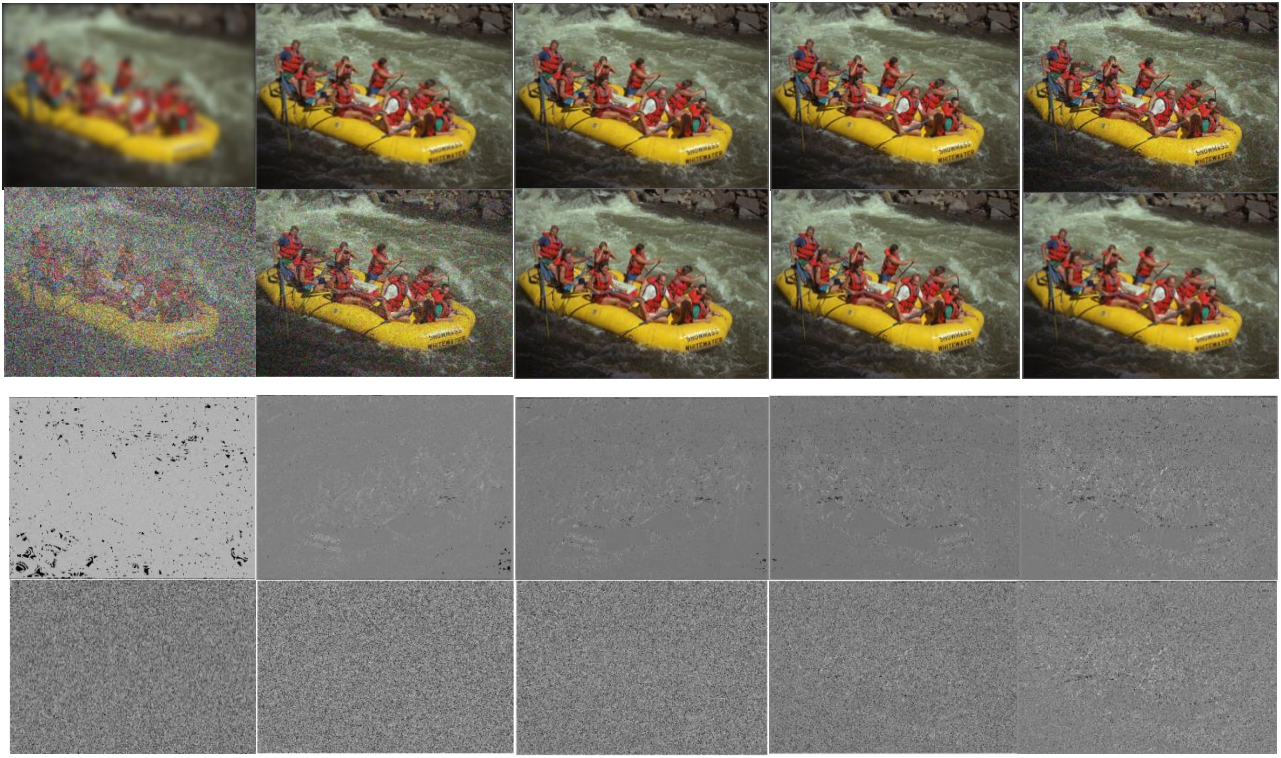

Fig.5 Ten degraded versions of an image sampled from the LIVE database [23], of which five images are blurred to five different degrees and five images are corrupted with five degrees of white noise. First and second rows contain blurred and white noise images respectively, in which the degree of image distortion in each row increases from right to left. Corresponding Stem Noise Energy Maps (SNEMs) of the images in the first and second rows are shown in the third and fourth rows, respectively.

For generality, let $a_{0\beta} = 1$ be the multiplier of $\hat{x}_\beta[n]$. In the case of a third-order model, squaring both sides of (5) and then taking the expectation (see appendix), of equation (12), the implicit stem noise energy in each image block is derived as follows:

$$b_{0\beta}^2 E\{i_\beta^2\} = R_{\hat{x}_\beta}(0)\left(\sum_{m=0}^{3} a_{m\beta}^2\right) + 2R_{\hat{x}_\beta}(1)\left(\sum_{m=0}^{2} a_{m\beta} a_{(m+1)\beta}\right) \quad (12)$$
$$+ 2R_{\hat{x}_\beta}(2)\left(\sum_{m=0}^{1} a_{m\beta} a_{(m+2)\beta}\right) + 2R_{\hat{x}_\beta}(3) a_{0\beta} a_{3\beta}$$

To visualize stem noise energy variations across the image blocks, a Stem Noise Energy Map (SNEM) can be created. The values of stem noise energy – computed in a block-based manner – are shown as greyscale pixels. Fig.5 shows an image chosen from the LIVE database [23] with ten degraded versions, of which five images are blurred with five levels of blurring and five images are corrupted with five levels of white noise. The first and second rows show blurred and white noise images respectively, in which the degree of image distortion in each row increases from right to left. Corresponding SNEMs of the images in the first and second rows are shown in the third and fourth rows. Note that as $2 \times 2$ pixels block have been used, the size of SNEMs are $1/4$ of original images. As expected, in a clockwise rotation the SNEMs are filled with more fine speckles. This phenomenon manifests as an expansion of the distribution (i.e., histogram) of stem noise energy. This is because white noise has a wide frequency spectrum and its autocorrelation function is very narrow. On the other hand, natural images have a decaying frequency spectrum and hence their autocorrelation function is wide. When images are blurred, higher frequencies are removed and the frequency spectrum gets progressively narrower. Conversely its autocorrelation function becomes wider. Thus, the degree of dispersion reflects image blurriness. Likewise, when images are degraded by more white noise, their frequency spectrum gets wider. This spread can be quantified via the standard deviation of the autocorrelation function, which is also the AC energy of the image.

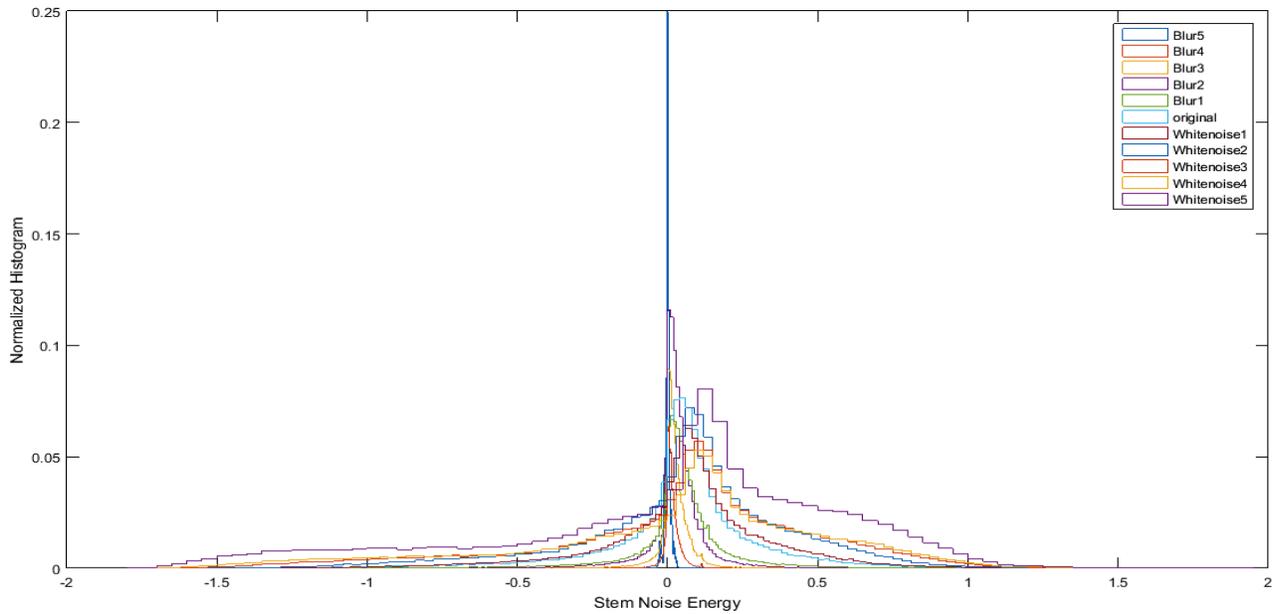

Fig.6 Normalized histograms of stem noise energy computed across the image blocks of ten images in Fig.4 and the relevant reference image.

Fig.6 shows normalized histograms of stem noise energy evaluated across the image blocks of ten images shown in Fig.5 and the appropriate reference image. Note that the values of the normalized input image i.e. $\hat{x}(u,v)$ may be negative (because the mean has been subtracted). As a result auto-correlation coefficients may also have negative values. Thus, negative values of stem noise energy computed from equation (12) are not unexpected. As it is evident from Fig.6, the images with more white noise degradation show a more dispersed stem noise energy histogram and conversely the images with more blurriness distortion show a less dispersed stem noise energy histogram. It is noteworthy that histogram of the original image is exactly in the middle. Fig.6 suggests that the first and second order statistics of the stem noise energy over blocks could be suitable quality-aware parameters, specifically for white noise and blurriness image distortions.

In contrast to white noise and blurriness image distortions – that alter the stem noise energy histogram– one of the characteristic attributes of the blockiness distortion includes peaks that are expressed as specific energy levels, originating from repetition of degraded blocks across the image. This blockiness distortion is a very common degradation resulting from the JPEG compression algorithm. As JPEG is a block-based coding scheme, it may lead to discontinuities at the block boundaries. Fig.7 illustrates noise energy histogram for a reference image (the same as reference image used in Fig.5) and its five blockiness degraded versions with varying severity. Images with more blockiness distortion and coarser blocks reveal themselves through more frequent and higher peaks. In accord with these findings, it will be seen later –in the experimental results section– that the mean and variance of stem noise energy do not form a suitable quality indicator for blockiness image distortion. In order to illustrate differentiating ability of computed noise energy for blockiness

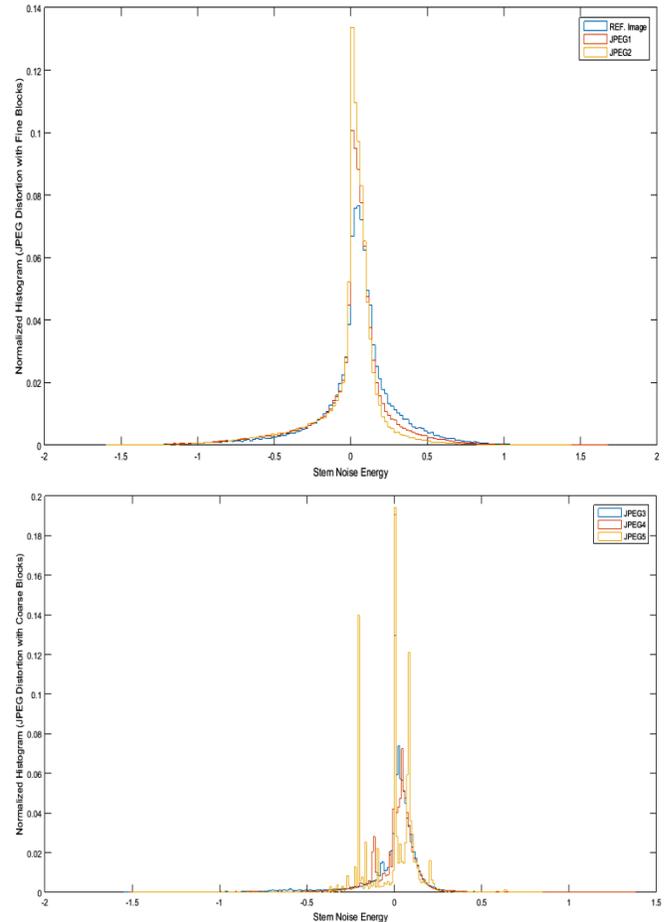

Fig. 7 Normalized histograms of stem noise energy computed for an image with blockiness distortion (a) With fine blocks (b) with coarse blocks

distortion, a multi-level thresholding is employed. Fig.8 shows an image resulted from applying a multi–level thresholding operation to the noise energy value computed from equation (12). Note that in this example the levels of thresholding were

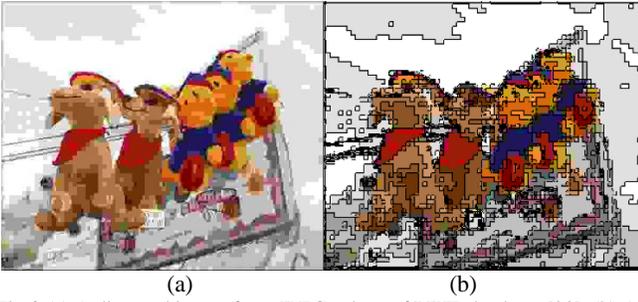

Fig.8 (a) A distorted image from JPEG subset of LIVE database [23]. (b) An image resulted from applying a simple thresholding for noise energy value (computed from (12)) for image (a).

optimized manually. As distorted image blocks own a definite range of energy emerged as specific peaks in the histogram, they could be segmented making use of such a thresholding operation.

Fig.9 shows the mean and variance of the stem noise energy computed for all the 494 distorted images in three canonical distortion subsets of the LIVE image database i.e. white noise, blurriness and JPEG. In Fig.9, each point is an image from the relevant image distortion subset. Blurriness simulates a loss of focus e.g. caused by a narrow depth of field in cameras, blockiness due to JPEG coding is inseparable from any compression scenario, e.g. for image storage and white noise could result from imaging in low light conditions, where sensor noise supervenes.

It is illustrated that blurriness, JPEG and white noise image distortions each occupy a different region in the stem noise energy space. In fact, there is a passage from blurriness to blockiness and then to white noise image distortion, in terms of the mean and variance of stem noise energy. This is compatible with image/video compression distortions, since as the codec's quantizer step size increases, the image initially becomes progressively blurred and then becomes blocky. Blurriness alleviates the effect of high frequencies and edges in the distorted image and white noise shows a contrary effect. While blurriness and white noise image distortions create low and high values of mean and variance of noise energy respectively, blockiness distortion depending on the size and number of degraded blocks and artificial patterns produce intermediate values. The importance of Fig.9 lies in the fact that it reveals the relationship between the most important signal-dependent and signal-independent image distortions – by capturing their characteristics with an independent, random component i.e. "Stem Noise" under a particular AR model.

Generally, in white noise distortion, images with greater degradations get greater values of mean and variance of noise energy and in the blurriness subset, increasingly blurred images show smaller values for mean and variance of noise energy. This is a result of the input of using the AR model; i.e. stem noise, is assumed *a priori* to be white noise. This kind of behavior is discussed in the next section with regard to human subjective ratings, suggesting that humans share the same kind of priors. In order to demonstrate the differentiation capabilities of noise energy characteristics, 2D-histograms of blurriness, blockiness and white noise distortions are shown in Fig.10. In Fig.10 the x and y axes are the mean and variance of stem noise energy respectively and z axis represents the number of degraded images.

In the experimental results section, Cardinal image distortions such as white noise, blurriness, blockiness, JPEG2000 and fast fading are considered. In addition, the experiments were performed using a wide range of well-known image databases, including the ones that contain multiple distorted images, with several distortions in play simultaneously.

### III. Experimental Results

Under the AR model described in the previous section, different image distortions shift the noise energy levels in a distinct fashion. This suggests that statistical characteristics of stem noise energy could be considered as the basis of a quantitative framework for image distortions. In the appendix the associated footprint of image distortions is provided for several image databases i.e. LIVEMD[24], MDID2013[25], TID2008[26],TID2013[27], MDID[28], QACS[29], CSIQ[30] andVDID2014[31].

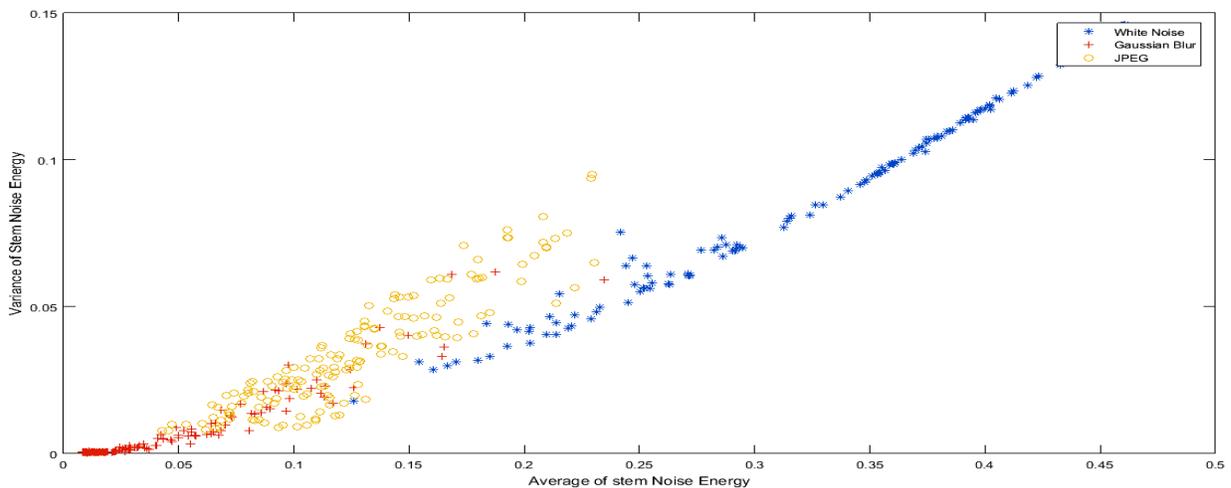

Fig.9 Scatter plot between average and variance of stem noise energy for distorted images in Gaussian blur, JPEG and white noise subsets of the LIVE [23]

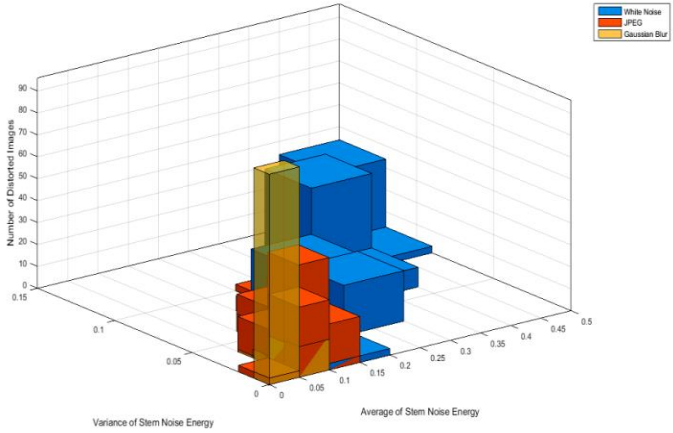

Fig.10 2D-histograms of blurriness, blockines and white noise image distortions in LIVE [23]. *x* and *y* axes are the average and variance of computed stem noise energy respectively. Z axis represents the number of degraded images in distortion subsets.

Herein, the performance of the proposed approach for quality assessment purposes was tested and evaluated on LIVE Image Quality Assessment Database [23], with regard to subjective quality scores assigned to each image by human users. The LIVE [23] database contains 29 reference images, each corrupted by several levels of five distortion types: Gaussian blur (Blur), JPEG compression, JP2K compression, Fast-Fading channel distortions (FF) and white noise (WN). JPEG2000 coding produces ringing near edges due to wavelet-based compression. Fast fading images in LIVE database are actually multi-distorted, first compressed into a bitstream using a JPEG2000 codec, then passed through a Rayleigh fast fading channel to simulate packet loss. The total number of distorted images is 779 and all distorted images are accompanied by their corresponding human subjective scores, i.e. Difference Mean Opinion Score (DMOS).

### A. Correlation with subjective scores

The predictive performance of models was evaluated by the Spearman Rank-Order Correlation Coefficient (SROCC), the Pearson Linear Correlation Coefficient (LCC) and the Root Mean Squared Error (RMSE). The SROCC measures prediction monotonicity, while the LCC and RMSE determine prediction accuracy.

In order to predictive validity of stem noise energy in image quality measurements, the Spearman Rank-Order Correlation Coefficient (SROCC) (defined in the following) was calculated between the statistics of stem noise energy and human DMOS scores, for all the distorted images in each distortion subset of the LIVE image database.

For N pairs of data as $(x_i, y_i)$, SROCC is computed as:

$$SROCC = \frac{\sum(X_i - X')(Y_i - Y')}{\sqrt{\sum(X_i - X')^2}\sqrt{\sum(Y_i - Y')^2}} \quad (13)$$

where $X_i$ and $Y_i$ are ranks of $x_i$ and $y_i$ respectively, $X'$ and $Y'$ are average ranks.

Note that the characteristics of stem noise energy are not used in any training-based frameworks or linear mapping functions, calculating prediction precision does not convey any special information, so RMSE and LCC are not reported. These parameters could be calculated for evaluating the performance of an image quality metric when using training strategy for mapping quality-aware features to subjective scores.

Nevertheless, without using any training procedure (or mapping approaches such as logistic functions, linear pooling etc. which are generally exploited by IQA metrics) the SROCC values show a substantial correlation with DMOS scores for white noise, blurriness, JPEG2000 and fast-fading image distortion subsets. Scatter plots between the mean of absolute values of stem noise energy and their DMOS values for white noise and blurriness image distortions are shown in Fig.11. As Fig.11 illustrates, white noise and blurriness image distortions displace $\bar{E}_{v_{(st)}}$(Mean of stem noise energy) computed for reference images. While for blockiness image distortion, such a displacement and also the correlation between $\bar{E}_{v_{(st)}}$ and DMOS values are not sufficient for quality measurement purposes, this is not the case in white noise and blurriness distortions in which there is a tight correlation between DMOS values and $\bar{E}_{v_{(st)}}$. As evident in the Gaussian blur subset, with any increase in the mean of stem noise energy, the subjective quality increases. This is in contrast to white noise image distortion, for which by any increase in the mean of stem noise energy, the subjective quality decreases.

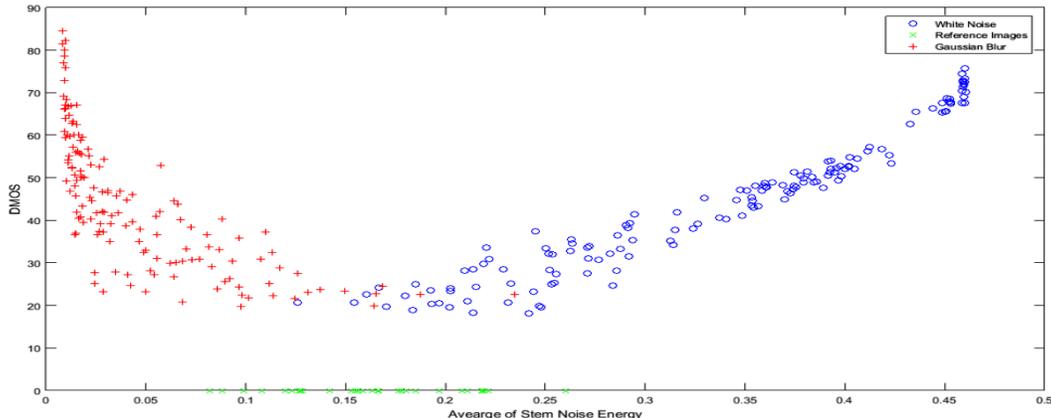

Fig.11 Scatter plot between average of absolute values of stem noise energy and DMOS values for white noise, Gaussian blur and reference images in the LIVE image database [23].

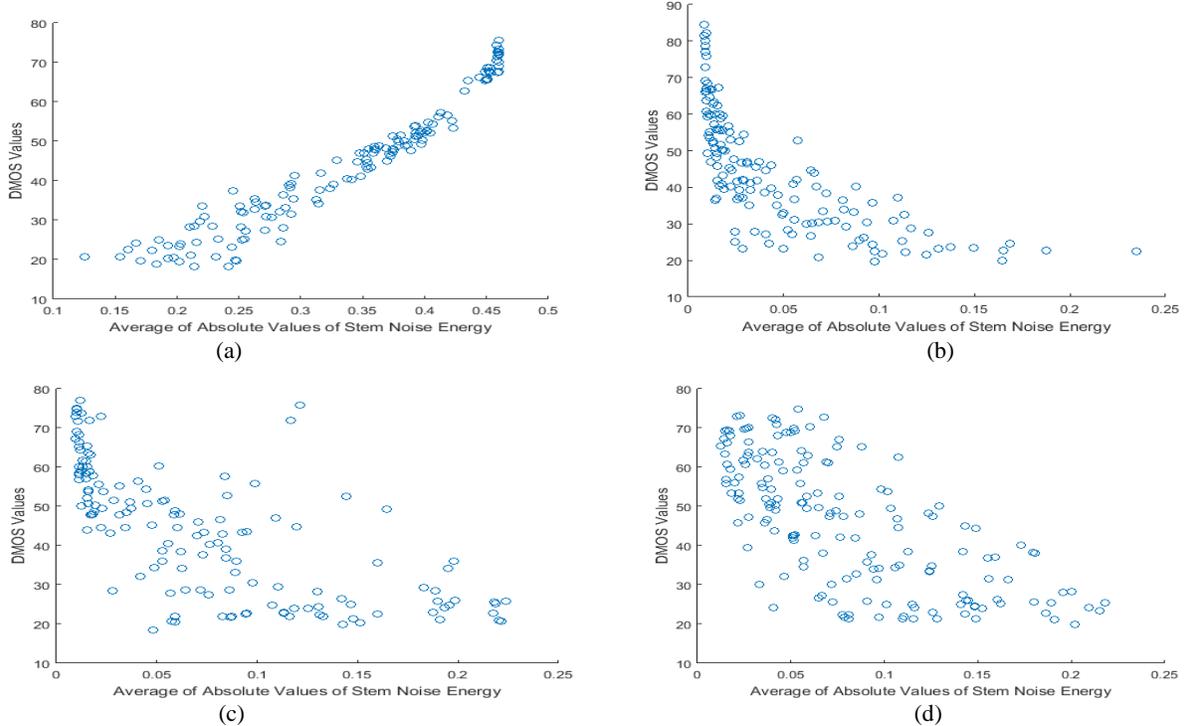

Fig.12 (a), (b), (c) and (d) scatter plots between average of absolute values of stem noise energy and DMOS values for white noise, Gaussian blur, fast-fading, and JPEG2000 subsets of the LIVE image database [23].

Fig.12 reports scatter plots between DMOS values and the mean of absolute values of stem noise energy computed for white noise, Gaussian blur, fast-fading and JPEG2000 subsets of the LIVE image database.

Table 1 reports the SROCC values between statistics of AR space of the pre-defined AR model and DMOS values for subsets of the LIVE [23] image database. This shows that stem noise energy outperforms all other statistics of the AR parameters for white noise, blurriness, JPEG2000 and fast fading distortions. For JPEG – due to its different subjective appearances depending on the size and location of the degraded blocks – $\bar{E}_{v_{(st)}}$ and $\sigma^2_{\bar{E}_{v_{(st)}}}$ have less predictive validity.

The efficiency of the proposed approach is also compared with other well-known methods in Table 2.

As NFEQM [5] is the basis of free-energy based approaches – that e.g. NFERM [6] surpass image quality metrics, such as BLIINDS [1], BRISQUE [2], DIIVINE [3] and C-DIIDINE [4] – it is included in our comparison as a benchmark of free-energy based quality assessment methods. In Table 2, the best results for each subset are highlighted in bold. In contrast to NFEQM [5] results –reported for white noise and blurriness distortions after employing a four-parameter logistic function for mapping to the human subjective scores– for the proposed approach the raw results were used. Table 2 demonstrates supremacy of the stem noise energy. The highest correlation in white noise subset belongs to the stem noise approach that reaches 0.9764. As the input of the AR model is white noise, high correlations with human scores in white noise subset is particularly reassuring.

Table I
SROCC values between statistics of AR space and DMOS values for subsets of the LIVE image database

| | White Noise | Gaussian Blur | JPEG2000 | Fast-Fading | JPEG |
|---|---|---|---|---|---|
| Mean of AR Coeffs. | +0.4346 | -0.2008 | -0.1014 | -0.3871 | +0.7212 |
| Variance of AR Coeffs. | +0.1179 | -0.7978 | +0.2420 | +0.5492 | +0.6494 |
| Mean of Horizontal AR Coeffs. | +0.1227 | -0.2256 | +0.0793 | -0.3664 | +0.5537 |
| Variance of Horizontal AR Coeffs. | -0.3383 | +0.0332 | +0.2729 | -0.1750 | +0.8394 |
| Mean of Vertical AR Coeffs. | +0.3985 | +0.1704 | -0.1347 | +0.0797 | -0.0559 |
| Variance of Vertical AR Coeffs. | +0.8417 | +0.3799 | -0.2053 | +0.2438 | -0.6170 |
| Mean of Main-Diagonal AR Coeffs. | -0.9158 | -0.7206 | -0.3805 | -0.7411 | -0.6827 |
| Variance of Main-Diagonal AR Coeffs. | -0.8867 | +0.3754 | +0.2461 | +0.4667 | -0.4780 |
| Mean of Secondary-Diagonal AR Coeffs. | +0.5527 | +0.1402 | -0.1030 | -0.0618 | -0.5271 |
| Variance of Secondary-Diagonal AR Coeffs. | -0.7294 | +0.5613 | -0.1032 | +0.2794 | -0.8667 |
| $\bar{E}_{v_{(st)}}$ (Mean of Stem Noise Energy) | **+0.9764** | **-0.8670** | **-0.7077** | **-0.7623** | -0.3012 |
| $\sigma^2_{\bar{E}_{v_{(st)}}}$ (Variance of Stem Noise Energy) | **+0.9623** | **-0.8146** | **-0.7023** | **-0.7416** | -0.4149 |
| $\bar{E}_{v_{(st)}}$ * | **+0.9691** | **-0.8638** | **-0.6767** | **-0.7651** | -0.1498 |
| $\sigma^2_{\bar{E}_{v_{(st)}}}$ * | **+0.9667** | **-0.9039** | **-0.7792** | **-0.7853** | -0.5824 |

*With full *R1* coefficient

Table II
SROCC values between mean of stem noise energy and DMOS values for WN and Gaussian blur subsets of the LIVE image database, in comparison with other methods

|  | White Noise | Gaussian Blur |
|---|---|---|
| NFEQM | 0.968 | 0.886 |
| Q Metric | 0.879 | 0.787 |
| JNBM | - | 0.549 |
| SINE | 0.957 | - |
| $\bar{E}_{v_{(st)}}$ (Mean of Stem Noise Energy) | **0.9764** | 0.8670 |
| $\sigma^2_{E_{v_{(st)}}}$ * (Variance of Stem Noise Energy) | 0.9667 | **0.9039** |

In other words, poor results for white noise image distortion indicates deficiency in model performance. Unlike NFEQM [5], NFERM [5] and all other AR-based image quality metrics which estimate AR parameters via least square, the current approach utilizes Yule-Walker equations, in which the auto-correlation function is embedded. The auto-correlation function reflects characteristics of different image distortions and contributes to the efficiency of the stem noise energy estimation.

## IV. CONCLUSION

While other no-reference metrics are predicted on some characteristics that are indirectly related to the amount of noise and distortion in degraded images, here the stem noise energy has been introduced as an explicit metric. The input of AR model that models a low energy version of the input image, has been called "Stem noise". In the current approach, the emphasis is on estimating stem noise energy in each block of distorted images. Beside NSS-based, No-reference IQA metrics that utilize GGD, Asymmetric GGD etc. fitting in spatial or transform (DCT,WAVELET etc.) domain, recently free-energy based methods have received attention. There are two major differences between the proposed approach and previous free-energy based methods. The first is while state-of-the art, free-energy based quality assessment methods are based on the systems parameters, i.e. AR parameters; here the hyperparameters describing the amplitude of random fluctuations (i.e., stem noise) have been used. It is noteworthy that in the latter case, the relevant AR model might not be necessarily optimal; from the viewpoint of residual error. While other approaches generally use least square methods, this paper uses the Yule-Walker equations. Taking the latter strategy, different types of correlation models could be incorporated in the modeling procedure. In the experimental results section, we have seen that there is a substantive correlation between characteristics of the stem noise energy and subjective human scores of image quality –for specific types of image distortions. This means that assessing the second order statistics under a model of how images are generated may not only be an important aspect of image reconstruction but may recapitulate the optimal processing shown in the human visual system.

# Appendix

- For generality, let $a_{0\beta} = 1$ be the multiplier of $\hat{x}_\beta[n]$. In the case of a third-order model, squaring both sides of AR model in equation (5) and then taking the expectation, yields:

$$\left( a_{0\beta}^2 \underbrace{E(\hat{x}_\beta^2[n])}_{R_{\hat{x}_\beta}(0)} + a_{1\beta}^2 \underbrace{E(\hat{x}_\beta^2[n-1])}_{R_{\hat{x}_\beta}(0)} + a_{2\beta}^2 \underbrace{E(\hat{x}_\beta^2[n-2])}_{R_{\hat{x}_\beta}(0)} + a_{3\beta}^2 \underbrace{E(\hat{x}_\beta^2[n-3])}_{R_{\hat{x}_\beta}(0)} \right)$$

$$+ 2\left( a_{0\beta}a_{1\beta} \underbrace{E(\hat{x}_\beta[n]\hat{x}_\beta[n-1])}_{R_{\hat{x}_\beta}(1)} + a_{1\beta}a_{2\beta} \underbrace{E(\hat{x}_\beta[n-1]\hat{x}_\beta[n-2])}_{R_{\hat{x}_\beta}(1)} + a_{2\beta}a_{3\beta} \underbrace{E(\hat{x}_\beta[n-2]\hat{x}_\beta[n-3])}_{R_{\hat{x}_\beta}(1)} \right)$$

$$+ 2\left( a_{0\beta}a_{2\beta} \underbrace{E(\hat{x}_\beta[n]\hat{x}_\beta[n-2])}_{R_{\hat{x}_\beta}(2)} + a_{1\beta}a_{3\beta} \underbrace{E(\hat{x}_\beta[n-1]\hat{x}_\beta[n-3])}_{R_{\hat{x}_\beta}(2)} \right) +$$

$$2\left( a_{0\beta}a_{3\beta} \underbrace{E(\hat{x}_\beta[n]\hat{x}_\beta[n-3])}_{R_{\hat{x}_\beta}(3)} \right) = b_{0\beta}^2 E(i_\beta^2[n])$$

$$R_{\hat{x}_\beta}(0)\left(\sum_{m=0}^{3} a_{m\beta}^2\right) + 2R_{\hat{x}_\beta}(1)\left(\sum_{m=0}^{2} a_{m\beta}a_{(m+1)\beta}\right) + 2R_{\hat{x}_\beta}(2)\left(\sum_{m=0}^{1} a_{m\beta}a_{(m+2)\beta}\right) + 2R_{\hat{x}_\beta}(3)a_{0\beta}a_{3\beta} = b_{0\beta}^2 E(i_\beta^2[n])$$

- Description of databases

 - LIVEMD (LIVE Multiply Distorted Image Quality Database)

This database contains images and results from a subjective study. The study was conducted in two parts. Part 1 deals with blur followed by JPEG, and part 2 with blur followed by noise. Each part includes 225 distorted images. The database consists of 15 reference images. In the relevant diagram, highest values of the stem noise energy are for images in the blurnoise subset which are contaminated with more white noise. In this subset, the images in which the dominant distortion is the blurriness, get lowest values of the stem noise energy. Likewise, in the blurjpeg subset, images in which the dominant distortion is blockiness due to the JPEG compression, get middle values of the stem noise energy. In this subset, images in which the dominant distortion is the blurriness, get lowest values of the stem noise energy.

-TID2008 (Tampere Image Database 2008)

The TID2008 contains 25 reference images and 1700 distorted images (25 reference images x 17 types of distortions x 4 levels of distortions). All images are saved in database in Bitmap format without any compression. The MOS was obtained from the results of 838 experiments carried out by observers from three countries: Finland, Italy, and Ukraine (251 experiments have been carried out in Finland, 150 in Italy, and 437 in Ukraine). In the relevant diagram; white noise, blockiness and blurriness image distortions along with the reference images have been considered. It is evident that the reference images and blockiness distortion occupy almost a common region, while white noise and blurriness distortions lie at the both ends of the diagram.

-CSIQ (Computational and Subjective Image Quality)

The database consists of 30 original images, each distorted using one of six types of distortions, each at four to five different levels of distortion. The CSIQ images were subjectively rated based on a linear displacement of the images across four calibrated LCD monitors placed side-by-side with equal viewing distance to the observer. The database contains 5000 subjective ratings from 35 different observers, and the ratings are reported in the form of DMOS. In the relevant diagram, just white noise, blurriness and blockiness image distortions along with the reference images have been considered. As it is evident, the reference images and images contaminated with the blockiness distortion occupy almost the same region in the stem noise space. The images degraded with white noise get higher values of the stem noise energy and images degraded with blurriness get lower values of the stem noise energy.

-MDID (Multiply Distorted Image Database)

MDID is an image database especially designed for evaluating the performance of image quality assessment algorithms on multiply distorted images. It contains 20 reference images and 1600 distorted images. Five distortions are introduced to obtain the distortion images, including:
Gaussian Noise, Gaussian Blur, Contrast Change, JPEG and JPEG2000. Each distorted image is derived from degrading the reference image with random types and random levels of distortions. Images with more Gaussian noise and more Gaussian blurr, occupy the highest and lowest stem noise energy at the two tails of the diagram respectively.

-TID2013(Tampere Image Database 2013)

The TID2008 contains 25 reference images and 3000 distorted images (25 reference images x 24 types of distortions x 5 levels of distortions). All images are saved in database in Bitmap format without any compression. The MOS was obtained from the results of 971 experiments carried out by observers from five countries: Finland, France, Italy, Ukraine and USA (116 experiments have been carried out in Finland, 72 in France, 80 in Italy, 602 in Ukraine, and 101 in USA). Again the reference images lie at the middle region of the diagram. Images with more white noise form one tail of the diagram with higher values of stem noise energy and images with more blurriness form another tail of the diagram with lower values of the stem noise energy.

-MDID2013 (Multiply Distorted Image Database)

Images in MDID2013 come from 12 pristine images. They span a wide range of scenes, colors, illumination levels and foreground/background configurations. In a practical image communication system, images usually undergo the stages of acquisition, compression and transmission, are presumably distorted with the artifacts of Gaussian blurring, JPEG compression and white noise injection in order. The overall 324 testing images are generated by successively corrupting each original image with blur, JPEG compression and noise. In this database, images contaminated with more white noise get higher values of stem noise energy and images contaminated with more blurriness distortion get lower stem noise energy values. Images with   JPEG compression get middle values.

-VDID2014 (Viewing Distance-changed Image Database)

This database consists of 160 images generated from eight pristine versions of two typical aspect ratios (height/width), and 320 differential MOS (DMOS) values collected from 20 inexperienced observers at two typical viewing distances, i.e. four and six times of the image height in terms of the ratio of the two physical distances. A total number of 160 images were produced by adding four commonly encountered distortion types: Gaussian blur, white noise, and JPEG2000 and JPEG compressions. While Gaussian blur and white noise distortions lie at both ends of the diagram with lower and higher values respectively, JPEG2000 and JPEG distortions lie at the middle of the diagram.

- Footprints of Image Distortions

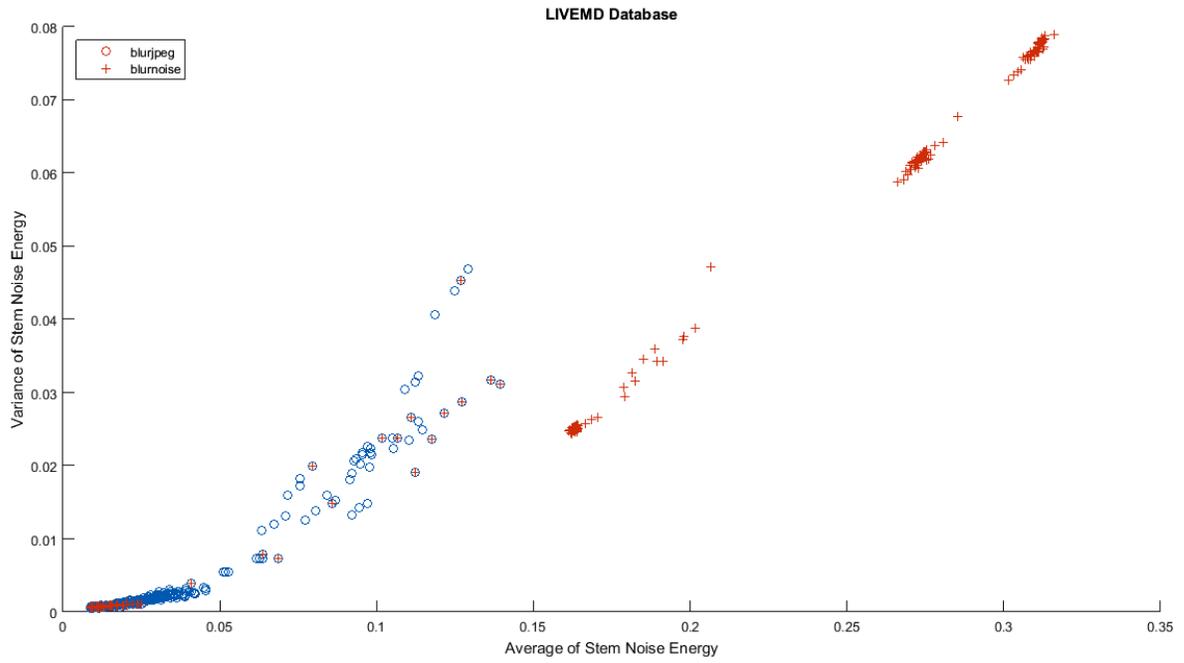

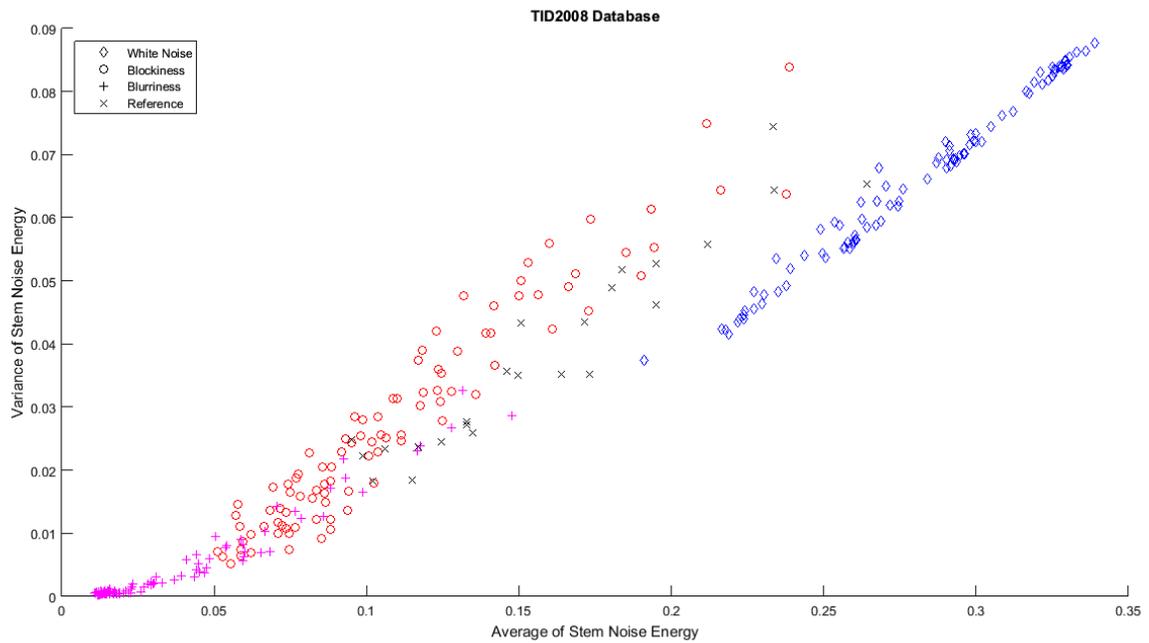

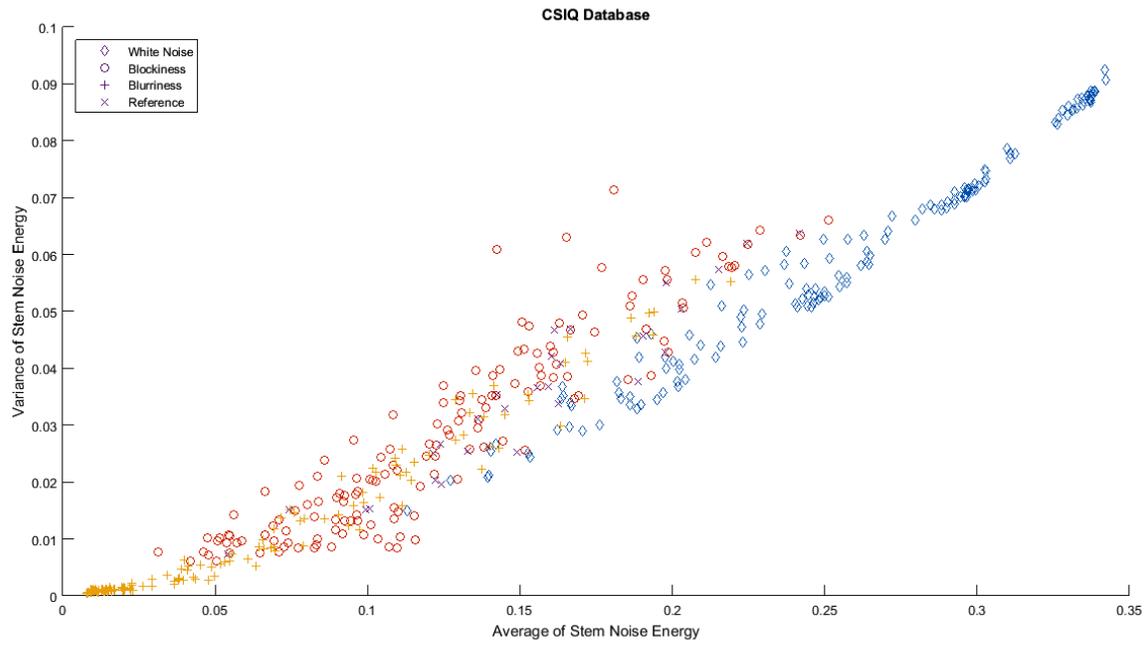

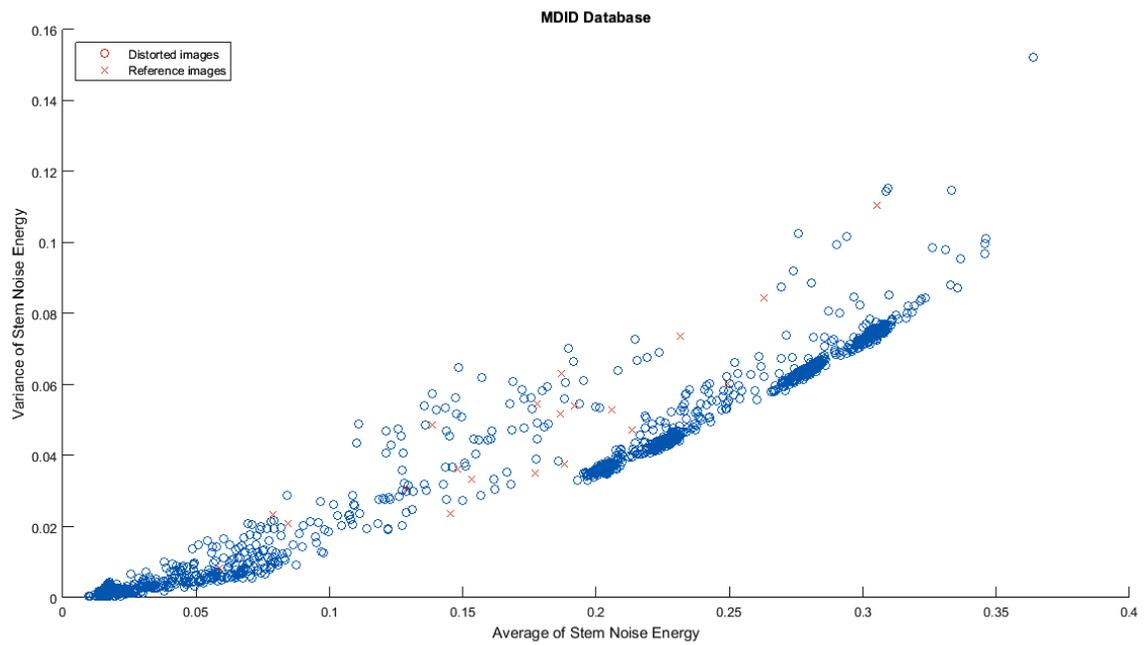

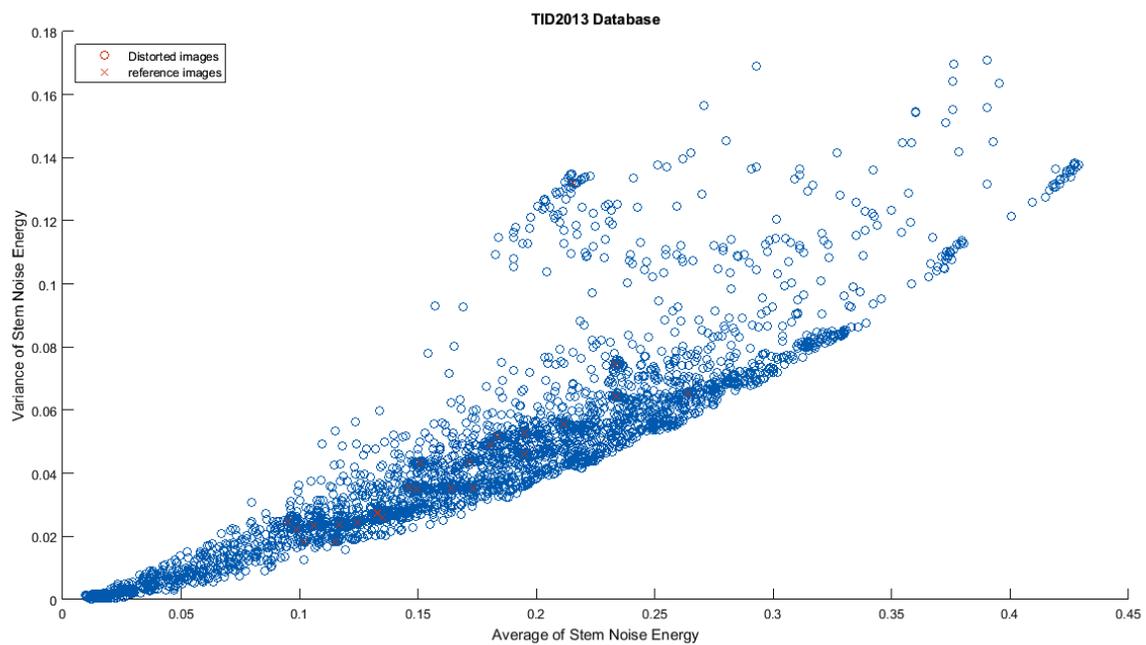
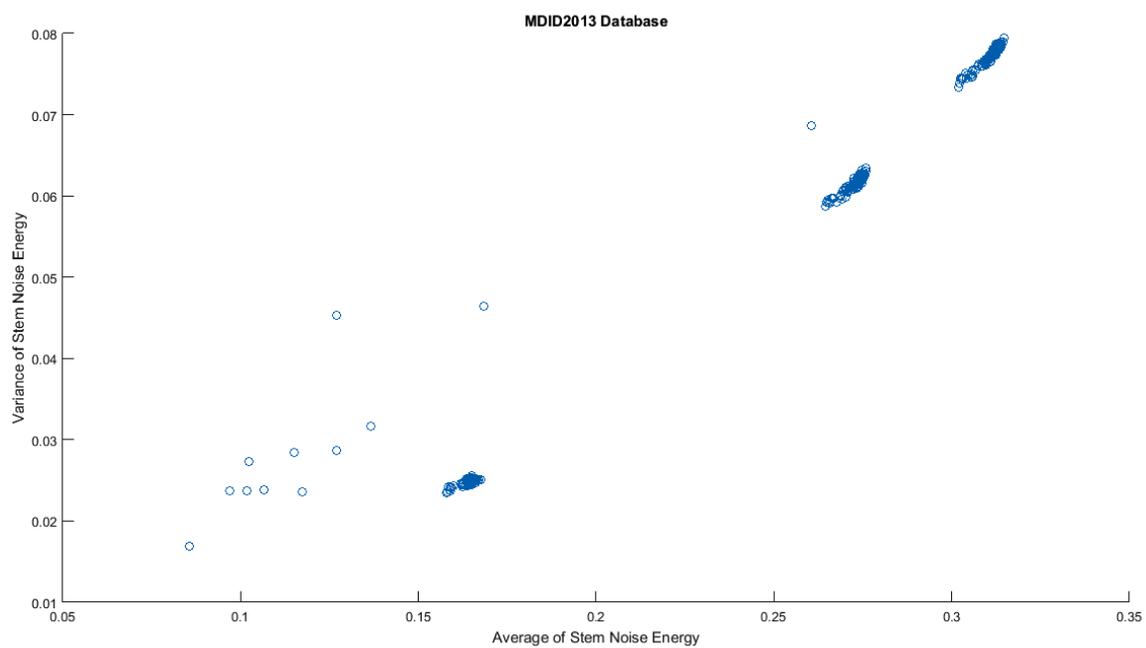

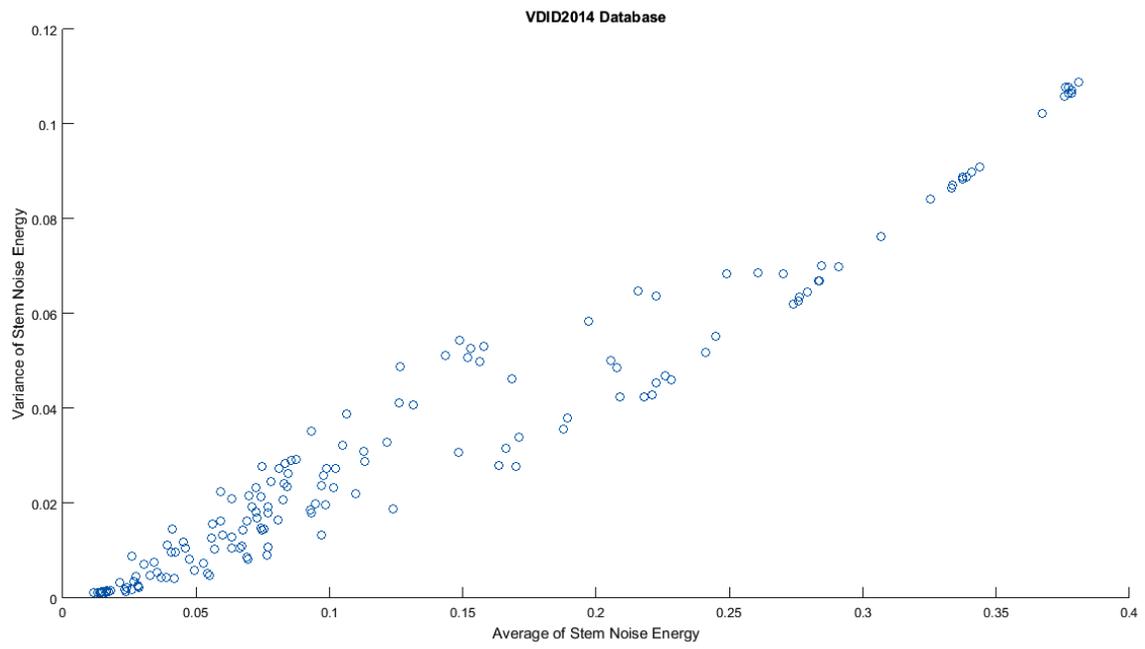